\documentclass[aps,prl, superscriptaddress,twocolumn]{revtex4-2}
\usepackage{graphicx}
\usepackage{dcolumn}
\usepackage{bm}
\usepackage{color}
\usepackage{siunitx}

\usepackage{mathtools}
\usepackage{hyperref}

\usepackage{ulem}
\allowdisplaybreaks

\begin{document}

\preprint{}

\title{Coherent subcycle optical shock from superluminal plasma wake}

\author{H. Peng}
\affiliation{Shenzhen Key Laboratory of  Ultraintense Laser and Advanced Material Technology, Center for Advanced Material Diagnostic Technology, and College of Engineering Physics, Shenzhen Technology University, Shenzhen 518118, China}

\author{T.W. Huang}
\email{taiwu.huang@sztu.edu.cn}
\affiliation{Shenzhen Key Laboratory of  Ultraintense Laser and Advanced Material Technology, Center for Advanced Material Diagnostic Technology, and College of Engineering Physics, Shenzhen Technology University, Shenzhen 518118, China}

\author{K. Jiang}
\affiliation{Shenzhen Key Laboratory of  Ultraintense Laser and Advanced Material Technology, Center for Advanced Material Diagnostic Technology, and College of Engineering Physics, Shenzhen Technology University, Shenzhen 518118, China}

\author{R. Li}
\affiliation{Shenzhen Key Laboratory of  Ultraintense Laser and Advanced Material Technology, Center for Advanced Material Diagnostic Technology, and College of Engineering Physics, Shenzhen Technology University, Shenzhen 518118, China}

\author{C.N. Wu}
\affiliation{Shenzhen Key Laboratory of  Ultraintense Laser and Advanced Material Technology, Center for Advanced Material Diagnostic Technology, and College of Engineering Physics, Shenzhen Technology University, Shenzhen 518118, China}

\author{M.Y. Yu}
\affiliation{Shenzhen Key Laboratory of  Ultraintense Laser and Advanced Material Technology, Center for Advanced Material Diagnostic Technology, and College of Engineering Physics, Shenzhen Technology University, Shenzhen 518118, China}

\author{C. Riconda}
\affiliation{LULI, Sorbonne Universit\'e, CNRS, \'Ecole Polytechnique, CEA, F-75252 Paris, France}

\author{S. Weber}
\affiliation{Extreme Light Infrastructure ERIC, ELI Beamlines Facility, 25241 Dolní Břežany, Czech Republic}

\author{C.T. Zhou}
\email{zcangtao@sztu.edu.cn}
\affiliation{Shenzhen Key Laboratory of  Ultraintense Laser and Advanced Material Technology, Center for Advanced Material Diagnostic Technology, and College of Engineering Physics, Shenzhen Technology University, Shenzhen 518118, China}

\author{S.C. Ruan}
\email{scruan@sztu.edu.cn}
\affiliation{Shenzhen Key Laboratory of  Ultraintense Laser and Advanced Material Technology, Center for Advanced Material Diagnostic Technology, and College of Engineering Physics, Shenzhen Technology University, Shenzhen 518118, China}

\date{\today}

\begin{abstract}
We propose exploiting the superluminal plasma wake for coherent Cherenkov radiation by injecting a relativistic electron beam (REB) into a plasma with a slowly-varying density up-ramp. Using three-dimensional particle-in-cell and far-field time-domain radiation simulations, we show that an ioslated subcycle pulse is coherently emitted towards the Cherenkov angle by bubble-sheath electons successively at the rear of the REB-induced superluminal plasma wake. A theoretical model based on a superluminal current dipole has been developed to interpret such coherent radiation, and agrees well with the simulation results. This radiation has ultra-short attosecond-scale duration and high intensity, and exhibits excellent directionality with ultra-low angular divergence and stable carrier envelope phase. Its intensity increases with the square of the propagation length and its central frequency can be easily tuned over a wide range, from the far-infrared to the ultra-violet.

\end{abstract}
\pacs{}

\maketitle

High-energy subcycle radiation pulses are useful for many applications, including attosecond-scale spectroscopy of living systems \cite{pupeza2020field}, ultrafast monitoring and control of molecules \cite{cocker2016tracking,peller2020sub}, attosecond metrology of electron motion, petahertz signal processing \cite{krausz2014attosecond,mucke2016pick,langer2018lightwave}, etc. Frequency-mixing \cite{steinleitner2022single,manzoni2015coherent}, pulse syntheses \cite{rossi2020sub,yang2021strong}, and parametric amplification \cite{lin2020optical}, etc. have been proposed for generating subcycle pulses, but these often need precise control and synchronization of the phases of the laser modes and the carrier envelopes. The pulse energy is also limited. In relativistic laser-plasma interaction studies, intense subcycle pulses can be generated from transverse focusing of an attosecond electron sheet injected into the laser wakefield by abrupt transition from vacuum to plateau\cite{Li2013Dense,li2014radially} and seed pulse amplification from interaction of a foil with an electron beam \cite{thiele2019electron} or laser wakefield \cite{simions2021laser}. However, these schemes either require a sharp density front or multiple driver beams. For many applications, it is desired to generate subcycle pulses with tunable frequency, stable carrier envelope phase (CEP), and foremost a simple setup.

\begin{figure}[htbp]
  \centering
  \includegraphics[width=0.482\textwidth]{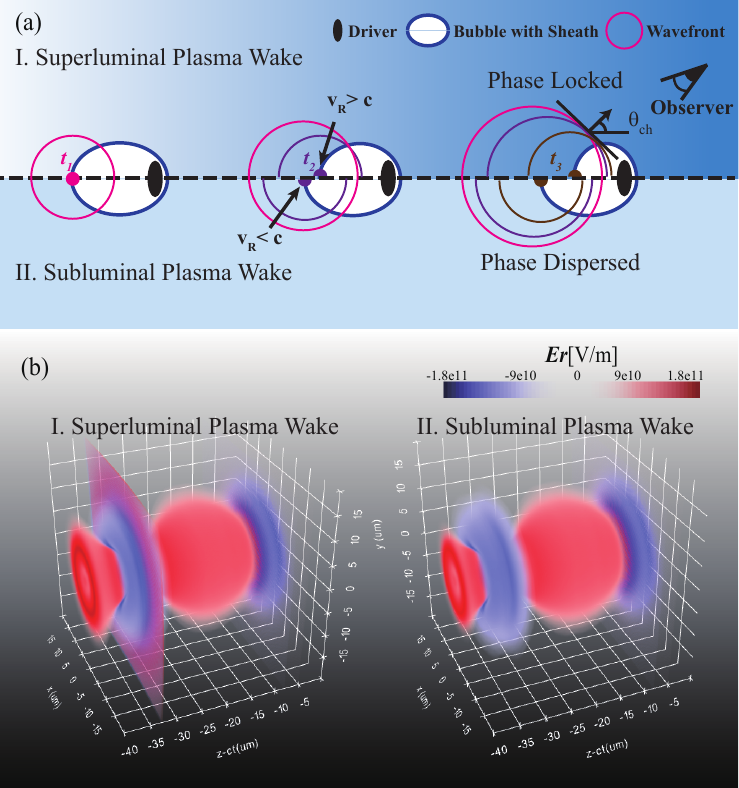}
  \caption{(a)Schematic diagram of the coherent optical shock generation from a superluminal plasma wake. All the spontaneous radiations are phase-locked at the Cherenkov angle in Case I, while they are phase-dispersed in Case II. Note the background color represents the plasma density, with darker colors indicating higher density. (b)3D PIC simulation results of subcycle pulse generation from a superluminal plasma wake at $z_0=\SI{1000}{\um}$, compared with that of a subluminal plasma wake, with no significant radiation generation. Shown is the radial electric field $E_r=E_x\cos\phi+E_y\sin\phi$ in a visualization of the subcycle pulse emitted from the ROB in Case I, where $\phi$ is the azimuthal angle and $\tan\phi=y/x$.}
  \label{fig:Schematic}
\end{figure}

It is well known that when an electron moves faster than the light in the medium, it emits Cherenkov radiation. The radiation emitted at different times is phase-locked and coherent at the Cherenkov angle\cite{cherenkov1934visible,tamm1937coherent}. No individual particles travel faster than light in vacuum and/or a plasma, while aggregates of particles can move superluminally and lead collectively to Cherenkov radiation\cite{bolotovskiui1972vavilov,vieira2021generalized}. In this Letter, we propose a simple scheme to generate coherent Cherenkov radiation by exploiting the rear of the bubble (ROB) driven by a REB in an underdense plasma with slowly-varying density up-ramp as superluminal radiation source, as demonstrated in Fig.\ref{fig:Schematic}. This new coherence mechanism\cite{vieira2021generalized} does not require bunching the radiating particles in a spatial region smaller than the radiation wavelength as that in existing coherent radiation sources\cite{li2014radially}.
Three-dimensional (3D) particle-in-cell (PIC) and far-field time-domain radiation simulations confirm the generation of an isolated subcycle pulse in the form of an optical shock in this regime. In particular, such radiation has many interesting and unique features. It is CEP stable, as well as of subcycle attosecond-scale duration. It has also excellent directionality, namely at the Cherenkov angle with very small angular divergence, and high intensity that scales with the square of the propagation distance. In addition, the frequency can be readily tuned by adjusting the densities of the plasma and REB.

As a high-density REB propagates into an underdense plasma, the coulomb force expels the electrons and forms a nonlinear wake bubble, the bubble size $r_b$ scales with the plasma density $n_e$ as $r_b\propto n_e^{-1/2}$\cite{xu2017high,tooley2017towards}. In a plasma with density up-ramp (down-ramp), the ROB moves forward (backward) relative to the driver in the comoving frame, as the bubble gradually shrinks (expands). The ROB velocity equals to the wake phase velocity, which is directly linked to the plasma density gradient along the propagation axis as\cite{Katsouleas1986Physical,xu2017high}:
\begin{gather}
v_R(z,t)=\frac{v_d}{1-(d\omega_p/dz)\omega_p^{-1}(v_dt-z)},
\label{eq:wakePhaseVelocity}
\end{gather}
where $v_d\lesssim c$ is driver beam velocity, $c$ is the speed of light in vacuum, $\omega_p(z)=\sqrt{e^2n_e(z)/\varepsilon_0m_e}$ is the plasma frequency corresponding to the local plasma density $n_e(z)$, $e$ and $m_e$ are the electron charge and mass, $\varepsilon_0$ is the vacuum permittivity, respectively. The ROB velocity can be superluminal ($v_R>v_d\sim c$) in a plasma with up-ramped density ($d\omega_p/dz>0$) and surpass the light phase velocity $v_{ph}\sim c$ in a tenuous plasma. While the ROB is subluminal ($v_R<v_d\sim c$) in a plasma with constant or down-ramped density ($d\omega_p/dz\le 0$). The sheath electrons flowing around the bubble emit strong spontaneous radiations at the ROB, where they have a large forward velocity and small curvature radius\cite{xu2017high}. The wavefronts of these spontaneous radiations travel with the speed of light outwards in a spherical shape, as shown in Fig.\ref{fig:Schematic}(a). The wavefronts of the spontaneous radiations from the superluminal ROB cross and form coherent radiation in the form of an optical shock at the Cherenkov angle $\theta_{ch}=\arccos{(v_{ph}/v_R)}$\cite{citedDiscussion1}, as shown in the upper panel of Fig.\ref{fig:Schematic}(a). For an observer at the Cherenkov angle(which is independent on the radiation frequency in the far field\cite{citedDiscussion1}), the arrival times of all wavefronts\cite{jackson1999classical} $t^{'}\approx t - \vec{n}\cdot\vec{r}_{R}/c+R/c=t(1-v_R\cos\theta/c)+R/c=R/c$ are constant, where $t$ is the time that the radiation is emitted from the ROB, $\vec{n}=(\sin\theta\cos\varphi, \sin\theta\sin\varphi, \cos\theta)$ is the propagation direction, $R$ is the distance from the origin to the observer and $\vec{r}_R=v_Rt\vec{e}_z$ is the trajectory of the ROB, respectively. While in a subluminal plasma wake, $v_R<v_{ph}$  and the spontaneous radiations emitted latter cannot catch up with those emitted earlier. The wavefronts are dispersed temporally and no coherent Cherenkov radiation, as shown in the lower panel of Fig.\ref{fig:Schematic}(a).

%%%%%%%%%%%%%%%%%%%%%%% 3D simulations ########################

The processes mentioned above are verified by 3D simulations using the WarpX PIC code\cite{myers2021porting}. To improve the accuracy and prevent numerical Cherenkov instabilities, the modified finite-difference Maxwell-equations solver CKC\cite{cowan2013generalized} as well as the pseudo-spectral Maxwell-equations solver PSATD are used\cite{blaclard2017pseudospectral,vincenti2018ultrahigh,vincenti2019achieving}, and they yield similar results. There are $512\times512$ cells in the transverse ($x,y$) directions and $1664$ cells in the axial ($z$) direction, with two electrons per cell. The immobile ions form a positive background. In the simulations, a moving window with speed $c$ is used. The size of the simulation box is $L_x=L_y=32c/\omega_{p0}$ and $L_z=13c/\omega_{p0}$, where $\omega_{p0}$ is the plasma frequency with density $n_0=\SI{1.5e18}{\per\cubic\cm}$. A complete blowout wake is generated by a high-current REB of density $n_b=16n_0$, whose normalized peak charge per unit length is $\Lambda=4\pi r_e\int_0^{\sigma_r}dr rn_b=4$ and the total charge is $Q\approx\SI{205}{\pico\coulomb}$, where $\sigma_r$ is the spot size of the beam and $r_e$ is the classical electron radius. The initial beam density is $n_b(r,\xi)=n_b$ for $r<\sigma_r$ and $z_l<z<z_r$, where $\xi=ct-z$, $\sigma_r=0.5{c}/{\omega_{p0}}$ and $\sigma_z=z_r-z_l=0.7{c}/{\omega_{p0}}$, respectively. The REB has a relativistic factor of $\gamma_b=2500$ and is assumed to have no energy spread. These parameters are similar to those of Ref. \cite{xu2017high}. Perfectly matched layers are implemented at all boundaries to avoid reflection of electromagnetic waves. Here we consider two cases. In Case I, a up-ramped plasma slab is used, the density from $z_i=\SI{500}{\um}$ to $z_e=\SI{2000}{\um}$ is $n_e(z)=n_0/[1-(z-z_i)/L_p]^2$, with $L_p\sim \SI{8174.2}{\um}$ (so that $n_e$ is $1.5n_0$ at $z_e$)\cite{citedDiscussion2}, and then sharply drops (within a distance $\SI{100}{\um}$) to vacuum. Thus, the velocity of the rear of the first bubble is $v_R\sim1.0021c$ along the up-ramp by substituting $v_dt-z$ with the bubble radius $r_m=2\sqrt{\Lambda}c/\omega_p$ in Eq.(\ref{eq:wakePhaseVelocity}). Such a plasma density profile has been used in previous simulations\cite{xu2016physics,Mehrling2017Mitigation} and can be realized in experiments by suitably tailoring gas capillaries\cite{SCHAPER2014Longitudinal}. In Case II, the plasma slab density is given by $n_e(z)=n_0=$constant in $\SI{100}{\um}<z<\SI{2100}{\um}$, with sharp drops (within $\SI{100}{\um}$) to vacuum on both ends. 

\begin{figure}[htbp]
  \centering
  \includegraphics[width=0.482\textwidth]{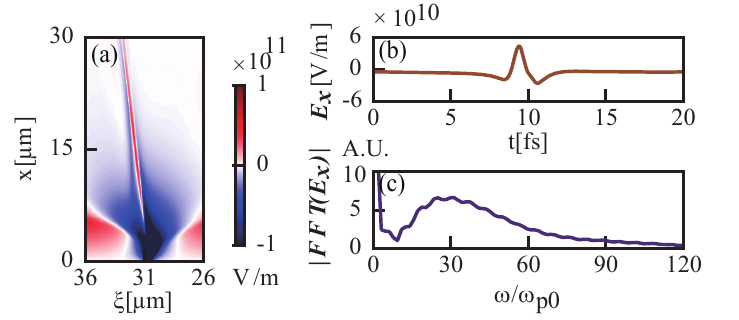}
\caption{ (a) Electric field  $E_x$ of the subcycle pulse from the ROB in the upper $x-\xi$ plane. (b) The temporal $E_x$ profile from (a) along the emission angle of $\SI{74.0}{\milli\radian}$. (c) Spectrum of the subcycle pulse in arbitrary units (A.U.). }
\label{fig:2Dsimulation}
\end{figure}

The radial electric field $E_r$ of the two cases is shown in Fig.\ref{fig:Schematic}(b). It is clearly seen that in Case I an isolated, subcycle and radially-polarized electromagnetic pulse in form of an optical shock emerges from the ROB. Such pulse is emitted successively all along the density up-ramp (see the Supplemental Movies\cite{SupplementalMaterial}). In Case II no such shock is generated.  Fig.\ref{fig:2Dsimulation}(a) for the subcycle radiation in Case I at $z_0=\SI{1000}{\um}$ shows that the emission angle is $\sim 74.0\pm 2.0~\si{\milli\radian}$. The duration of the pulse is $\sim\SI{581.4}{\as}$ (FWHM) and the field peak value is $\SI[per-mode=symbol]{44.1}{\giga\V\per\m}$, the same order as that of the wakefield, as shown in Fig.\ref{fig:2Dsimulation}(b). As expected, the corresponding spectrum, shown in Fig.\ref{fig:2Dsimulation}(c), is broad with the frequency bandwith $\Delta \omega\sim 60\omega_{p0}$. The frequency at its maximum strength is $\sim 31.3\omega_{p0}$, or wavelength $\sim \SI{870.7}{\nm}$. Note that the rear of the secondary bubbles can also emit subcycle radiation pulses. However, they are emitted with larger Cherenkov angles and are much weaker than that from the first bubble, see the details in the supplemental material(SM)\cite{SupplementalMaterial}.

%%%%%%%%%%%%%%%%%%%Electron Trajectories Analyses%%%%%%%%%%%%%%%%%%%%%%%

\begin{figure}[htbp]
  \includegraphics[width=0.482\textwidth]{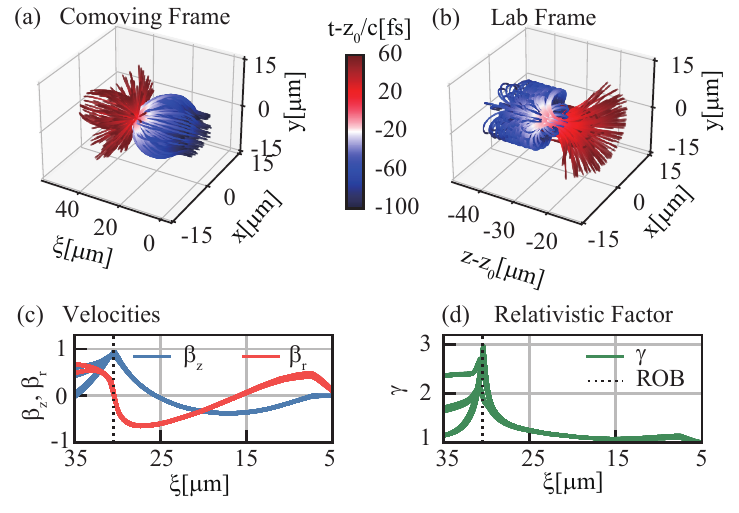}
  \caption{Trajectories of the sheath electrons passing through the ROB at $t=z_0/c$ in the (a) comoving and (b) laboratory frames. Their axial and radial velocities as well as their relativistic factors are shown in (c) and (d), respectively, where the dotted lines mark the position of the ROB at $\xi=ct-z\sim \SI{30.5}{\um}$. Note that the variations of the velocity and energy of almost all sheath electrons overlap before they reach the ROB.}
  \label{fig:sheathElectron_traj}
\end{figure}

To understand how the electron dynamics lead to subcycle radiation at the ROB, it is of interest to look at the trajectories of the sheath electrons and compute the far-field radiation based on these trajectories. In this case, simulations are carried out using the quasi-3D cylindrical-geometry FBPIC code\cite{lehe2016spectral}, with 416 cells along $z$, 12 particles per cell while all other parameters remain the same. Although the longitudinal spatial resolution $dz$ is reduced, it is still sufficient to resolve the electron dynamics involved. FBPIC can mitigate the computing resource consumption of large-scale plasma wake simulations\cite{wang2021free,shalloo2020automation,ke2021near},  and significantly reduces the number of electrons to be analyzed, which in turn reduces the computational burden of far-field radiation simulations as will be discussed next. Fig.\ref{fig:sheathElectron_traj} shows the trajectories and dynamics of about 400 sheath electrons of Case I passing through the ROB at $t=z_0/c$ in the co-moving and lab frames, respectively. The trajectories are mainly 2D in the $r-z$ plane and they all follow the same pattern: starting at about $r_i\sim\SI{5.3}{\um}$ and form the narrow sheath defining the bubble boundary, with maximum radius $r_m\sim\SI{8.1}{\um}$\cite{lu2006nonlinearPOP,lu2006nonlinearPRL}. As the REB propagates forward, the sheath electrons are eventually pulled back by the charge separation field of the nearly immobile plasma ions in the channel, merge at the ROB and get reflected, forming the bubble. At the ROB they have high (almost $c$) forward axial velocities but nearly zero radial velocities, with their relativistic factors reaching $\gamma_m\sim3.0$, as shown in Fig.\ref{fig:sheathElectron_traj}(c-d). So, at the ROB the electron trajectories have small curvature radius $\rho$ and the electrons emit subcycle bursts spontaneously contained within a cone of opening angle $\Delta\theta\sim1/\gamma$ centered on axis and with short duration $\tau_b\sim 1/\omega_c\sim\rho/\gamma^3c$, where $\omega_c$ is the critical radiation frequency\cite{jackson1999classical,corde2013femtosecond}(see details in the SM\cite{SupplementalMaterial}). 

While for the coherent Cherenkov radiation, the collective effect dominates over the single electron radiation. A simplified theoretical model is constructed (see details in the SM\cite{SupplementalMaterial}), in which the ROB is taken as a superluminal current dipole based on the electron dynamics near the ROB:
\begin{gather}
  \vec{J}(z, t)=-en_R\vec{v}_R=-en_Rc(\beta_{zm}\vec{e}_z+i\beta_{rm}\vec{e}_r)e^{i(\omega_{\beta}t-k_{\beta}z)},\label{eq:dipoleCurrent}
  \end{gather}
where $n_R=n_{max}\exp[-(z-v_Rt)^2/\Delta^2]$ is the density distribution of electron sheath at the ROB with width $\Delta$ and $v_R=\omega_{\beta}/k_{\beta}$, $n_{max}$ is the peak density, $\beta_{zm}$ and $\beta_{rm}$ are the maximum oscillation velocities in the longitudinal and radial direction (which are assumed as constant since they change very litte with a slowly-varing density up-ramp\cite{citedDiscussion2} and note they have a phase diffrence of $\pi/2$), $\omega_{\beta}$ and $k_{\beta}$ are the frequency and wavevector of the nonlinear plasma wake, respectively. The spectral intensity of the far-field radiation from this superluminal current dipole is\cite{landau2013classical}:
\begin{align}
  \frac{d^2 W}{d\omega d\Omega}=\frac{cR^2}{\pi\mu_0}|i\vec{k}\times \vec{A}_{\omega}(R)|^2=\frac{cR^2}{\pi\mu_0}|kA_{\omega}(R)\sin\theta|^2,
\end{align}
where $\vec{A}_{\omega}(R)=\frac{\mu_0}{4\pi}\frac{e^{ikR}}{R}\vec{J}(\omega, \vec{k})$ is the radiation vector potential and $\mu_0$ is the vacuum permeability. Then we have
$d^2 W/d\omega d\Omega\propto |\vec{J}(\omega, \vec{k})\sin\theta|^2\propto (I_sI_t\sin\theta)^2$, where $I_s$($I_t$) is the spatial (temporal) Fourier transform of $\vec{J}(z, t)$ and $\theta$ is the radiation angle. The scaling with $|\sin\theta|^2\sim|\theta|^2$ is different from that of the radiation of a single sheath electron which is peaked on axis and allows to improve the radiation intensity by changing the density gradient\cite{SupplementalMaterial}. $I_s$ can be written as:
\begin{gather}
  I_s=\int_{-\infty}^{\infty}e^{iz(\omega/v_R-k\cos\theta)}dz = \delta(\omega/v_R-k\cos\theta),
\end{gather}
which means the radiation is only emitted at the Cherenkov angle. Integrating from $0$ to $L_u$, we have $|I_s|=L_u$ at $\theta_{ch}$, i.e. the radiation intensity is proportional to the square of the propagation length, which is a typical feature of the coherent radiation. The spectral distribution is determined by the temporal integral of the Fourier transform:
\begin{gather}
I_t = \int_{-\infty}^{\infty}\exp\left (-\frac{v_R^2\tau^2}{\Delta^2}\right )e^{-i(\omega-\omega_{\beta})\tau}d\tau=\sqrt{2\pi}e^{\frac{-(\omega-\omega_{\beta})^2}{\omega_d^2}},
\end{gather}
where $\tau=t-z/v_R$, $\omega_d=2v_R/\Delta\approx 2c/\Delta$. The frequency bandwidth and the duration of the subcycle radiation are determined by $\Delta$. Since $c/\Delta\gg \omega_{\beta}$, $\omega_{\beta}$ has little effect on the spectral distribution. The sheath width at the ROB cannot be determined analytically\cite{lu2006nonlinearPOP,lu2006nonlinearPRL,golovanov2023energy}(but it can be tuned by changing the plasma density or driver strength\cite{SupplementalMaterial}), in the PIC simulation it is measured to be $\Delta\approx \SI{270.3}{\nm}$. The frequency bandwith is estimated to be $\Delta \omega^{'}\approx 2\omega_d\approx 64\omega_{p0}$, close to the simulation results. 

%%%%%%%%%%%%%%%% Radiation collected on a far-field detector plane%%%%%%%%%%%%%%%%

\begin{figure}[htbp]
  \centering
  \includegraphics[width=0.482\textwidth]{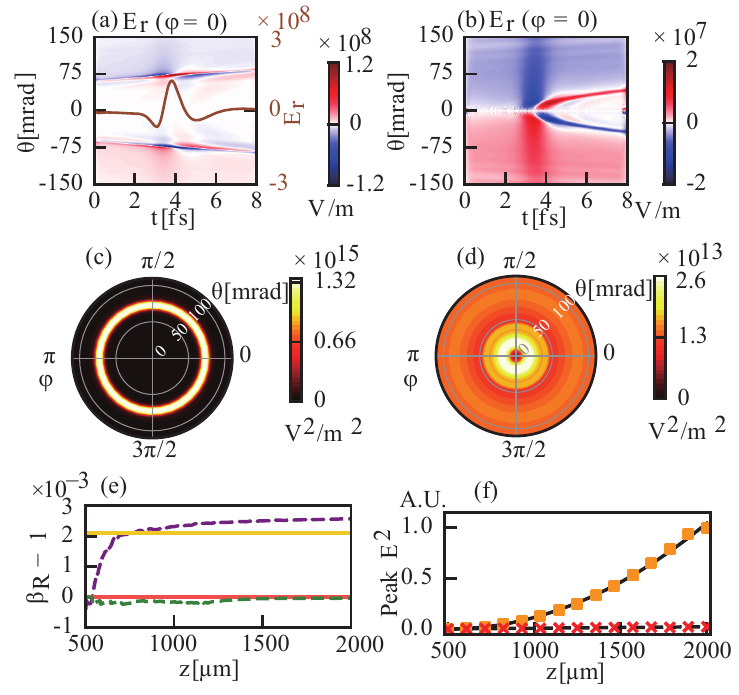}
  \caption{Far-field radiation computation results: (a,b) time-resolved radiation $E_r$ on the azimuthal plane $\varphi=0$  and (c,d) time-averaged radiation $\overline{\left\langle E^2 \right\rangle}=\int E^2(t)dt/\int dt$ on the $\theta-\varphi$ plane. (a,c) are from Case I  and (b,d) are from Case II, respectively. The brown superimposed line in (a) is the 1D slice of 2D $E_r$ at $\theta=\SI{72.7}{\milli\radian}$, at which angle $E_r$ obtains its maximum amplitude. It is labelled on the right axis. (e) Phase velocities of the first bubbles. Broken lines are for Case I (purple) and Case II (green), obtained by tracing the first ROB in FBPIC simulations. Solid lines are the theoretical values of Case I (yellow): $\beta_R=v_R/c=1.0021$ and of Case II (red) $\beta_R=1$. (f)Peak  radiation intensity $E^2$ in arbitrary units versus the propagation distance for Case I (orange squares) and Case II (red crosses). The solid line and broken line are the quadric and linear fit for the squares and crosses, respectively. }
  \label{fig:farFieldRadiation}
\end{figure}

A postprocessing far-field time-domain code FaTiDo\cite{SupplementalMaterial,pardal2023radio} was developed to compute the radiation based on the trajectories of the bulk electrons, as projected on a far-field spherical surface.  FaTiDo reads the trajectories of the macro-electrons and computes the total radiation field emitted by the bulk electrons. A far-field spherical detector plane is set \SI{1}{\m} away from the origin, i.e. $R=|\vec{r}_{obs}|=\SI{1}{\m}$. The detector time axis is defined to completely cover the arrival time of the radiation from the ROB. The time resolution is $dt_f=\SI{5e-18}{\s}=\SI{5}{\as}$ with $N_t=1600$ time steps. The spherical detector $\theta-\varphi$ plane is resolved as $N_{\theta}\times N_{\varphi}$ observers, with $N_{\theta}=128$ along $\theta$, $N_{\varphi}=32$ along $\varphi$ and $\theta$ ranging from $0$ to \SI{150}{\milli\radian}, $\varphi$ from 0 to $2\pi$, respectively. Note that computing far-field radiation by tracing macro-particles with large weight will greatly exaggerate the amplitude of incoherent radiation, but not for the coherent radiation \cite{pausch2018quantitatively}. 

The time-resolved as well as time-averaged far-field radiations are shown in Fig.\ref{fig:farFieldRadiation}. For Case I, the radiations along the driver path are coherently phase-locked at the Cherenkov angle $\theta\sim\SI{72.7}{\milli\radian}$, very close to the one observed in near-field PIC simulation and the one predicted by theory $\theta_{ch}=\arccos{(c/v_R)} \sim\SI{65.2}{\milli\radian}$. A subcycle pulse is clearly seen in Fig.\ref{fig:farFieldRadiation}(a), where the $E_x$ ($E_r$ at $\varphi=0$) temporal profile is almost the same as that in the near-field PIC simulation, indicating that the raidation pulse is CEP stable, and the duration $\sim\SI{617.1}{\as}$ (FWHM) is slightly longer. The time-integrated radiation is concentrated around the Cherenkov angle and forms a photon ring with very small angular divergence $\sim\SI{5.7}{\milli\radian}$ at FWHM. This is due to the fact that the superluminal phase velocity of the wakefield stays relatively constant, as can be seen in Fig.\ref{fig:farFieldRadiation}(e). Without the density ramp (Case II), the radiation is not phase-locked and incoherent, resulting in a much smaller amplitude and intensity, as shown in Figs.\ref{fig:farFieldRadiation}(b) and (d).

As shown in Fig.\ref{fig:farFieldRadiation}(f), in Case I the peak radiation intensity presents a quadric dependence on the propagation distance, while the dependence is linear in Case II. These are signatures of superradiant radiation and incoherent radiation\cite{vieira2021generalized} respectively, which agrees well with our theoretical model. The total energy of the subcycle pulse of Case I is $\epsilon=\int\vec{S}\cdot d\vec{\sigma}dt\approx \SI{11.5}{\micro\J}$, where $\vec{S}=(\vec{E}\times \vec{B})/\mu_0$ is the Poynting vector, $d\vec{\sigma}=R^2\vec{n}\sin\theta\cos\theta d\theta d\varphi$ is the surface element vector of the spherical detector and $\vec{B}=(\vec{n}\times\vec{E})/c$\cite{jackson1999classical}. Thus, the conversion efficiency is $\eta=\epsilon/\epsilon_b\approx \num{5.2e-5}$, where $\epsilon_b$ is the energy of the injected driver beam. Note that the pulse energy and conversion efficiency can be greatly improved by using a longer plasma slab while preserving the same density gradient as they scale with the square of the plasma length. Plasma slab length of the order of \SI{10}{\cm} can already be realized in experiments\cite{gonsalves2019petawatt}.

%%%%%%%%%%%%%%%% Discussions and Conclusions%%%%%%%%%%%%%%%%%%%%%%%%%

The proposed scheme is highly robust over a wide range of parameters and allows us to control the properties of subcycle pulse, e.g. pulse width and pulse intensity, which is supported by parameter scan over density profile shapes, density gradients, driver strengths (see details in the SM\cite{SupplementalMaterial}). Most importantly, it is possible to tune the central wavelength from the far-infrared to the ultra-violet by simply adjusting the plasma and beam densities while maintaining the same density profile shapes.

In summary, we have proposed a new scheme of exploiting the Cherenkov radiation for generating coherent isolated, intense, CEP-stable subcycle radiation pulses from a superluminal plasma wake in a plasma with density up-ramp. A theoretical model has been developed and it agrees well with the simulation results. In particular, the far-field radiation has an excellent directionality, low angular divergence, a well-defined wavefront and high frequency tunability. These attributes make the proposed method highly attractive for a variety of applications. It is worth noting that the necessary plasma and high-energy drivers are already available in current experimental setups\cite{nie2020photon,wu2021high}.

This work is supported by the National Natural Science Foundation of China (Grants No.12005148, No.12175154), the National Key R\&D Program of China (Grant No. 2022YFA1603300), the Shenzhen Science and Technology Program (Grant No. RCYX20221008092851073) and Foundation of Science and Technology on Plasma Physics Laboratory(Grants No.6142A04200211), the Natural Science Foundation of Top Talent of Shenzhen Technology University (Grant No. 2019010801001). H.P. acknowledges H. Zhang, C.X. Zhu for discussions on code implementation. S.W. acknowledges support from the projects High Field Initiative (CZ.02.1.01/0.0/0.0/15\_003/0000449) (HiFI)  and Advanced research using high intensity laserproduced photons and particles (ADONIS)(CZ.02.1.01/0.0/0.0/16\_019/0000789), both from European Regional Development Fund.


\begin{thebibliography}{41}
\bibitem [{\citenamefont {}()}]{citedDiscussion1}%
    \BibitemOpen
    \bibfield  {title} {\bibinfo {title} {As disucssed later, the emitted pulse is ultrashort with wide bandwith and most the frequency components $\gg\omega_p$. Moreover, it undergoes an additional refraction when exits the plasma and all the frequency arrive at the far-field detector at the same Cherenkov angle. Thus we can neglect the effect of the plasma permittivity and take $v_{ph}\sim c$}}\BibitemShut {NoStop}%
\end{thebibliography}
\end{document}